# Asymptotic solution of the Schrödinger equation for the elliptic wire in the magnetic field

Igor Bejenari[a] and Valeriu Kantser[b]

Institute of Electronic Engineering & Industrial Technologies, Academiei str., 3/3, MD2028 Chisinau, Moldova

The asymptotic solution of the Schrodinger equation with non-separable variables is obtained for a particle confined to an infinite elliptic cylinder potential well under applied uniform longitudinal magnetic field. Using standard-problem method, dimension quantized eigenvalues have been calculated when the magnetic length is large enough in comparison with the half of the distance between the boundary ellipse focuses. In semi-classical approximation, the confined electron (hole) states are divided into the Boundary States (BS), Ring States (RS), Hyperbolic Caustic States (HCS), and Harmonic Oscillator States (HOS). For large angular momentum quantum numbers and small radial quantum numbers, the BS and RS are grouped into the "Whispering Gallery" mode. They associate with particles moving along the wire cross section boundary. The motion is limited from the wire core by the elliptic caustic. Consisting of the HCS and HOS, the "Jumping Ball" modes correspond to the states of particle moving along a wire diameter when the angular momentum quantum number is much less than the radial quantum number. In this case, the motion is restricted by the hyperbolic caustics and two boundary ellipse arcs. For excited hole states in Bi wire, the energy spectrum and space probability distribution are analyzed.

---

[a] Email: bejenari@iieti.asm.md

[b] Email: kantser@iieti.asm.md

# 1. INTRODUCTION

The study of the factor of a carrier mass anisotropy in the thermoelectric and magnetotransport phenomena has attracted a big interest[1 2]. The most promising thermoelectric materials are multi valley IV-VI semiconductor compounds and bismuth – like semimetals with anisotropic effective mass carrier parameters[3 4]. The given paper presents mathematical background to study magnetic quantum oscillations of nanowire electronic characteristics in weak magnetic field, the carrier mass anisotropy being considered. We consider the elliptic cylinder quantum well model because the problem of size quantization of a carrier, with anisotropic effective mass parameters, in a cylindrical well is equivalent to that of a carrier with some isotropic effective mass in an elliptical well[5].

Subject: We present the solution of the Schrodinger equation in the elliptic domain when the longitudinal magnetic field is applied. We do not consider the free motion of particles along the wire. The Schrodinger eqution presents ordinary partial differential equation of the 2$^{nd}$ order with non-separable variables. Firstly, Nedorezov considered this equation using perturbation theory[6]. He obtained the eigenvalues corresponding to the Ring state and Boundary State eigenvalues from our theory.

Method. To solve the Schrodinger equation we apply Standard-Problem Method (etalon problem method)[7]. Initially, we reduce our equation to the one with known solution. In the magnetic field absence, the solution of the Schrodinger equation in the elliptic domain has been obtained earlier[8 9]. The solution of the simplified equation is used to elaborate the trial function. In our case, the trial function presents product of the exponent function and either Airy function or parabolic cylinder function. The argument of the functions is asymptotic expansion in terms of $\omega^{-1/3}$ (or $\omega^{-1/2}$), where $\omega$ is a large parameter (eigenvalue). The coefficients in front of parameter $\omega$ are unknown functions of two independent space variables. Then, we substitute the trial function in the original equation. Equating the corresponding coefficients in front of powers $\omega^{-r}$, we obtain a set of recurrent differential equations. The unknown functions are obtained as a solution of this set of equations. Applying the boundary condition, we calculate eigenvalues of the Schrödinger equation.

We describe the electron (hole) motion in the wire cross section by using caustics. This term is widely used in optics and theory of quantum resonators [9 10 11 12 13]. A caustic is a locus where the rays of geometrical optics have an envelope; at a caustic the amplitude has a singularity and the asymptotic expansion of geometrical optics is not valid. Kravtsov[14] and Ludwig[15] independently arrived at a means of overcoming this difficulty by introducing the asymptotic solution involving the Airy function. This solution is reduced to the geometrical optics solution on one side of the caustic. It is exponentially damped on the other side of the caustic. The solution remains finite at the caustic.

Coverage: Two main approximations are used in the paper. Firstly, we use semi-classical approximation valid for highly excited states. Secondly, the applied magnetic field is supposed to be small enough, so that the magnetic length, $L_B = \sqrt{\hbar/eB}$, is greater than the half of the distance between the boundary ellipse focuses $c$

$$c/L_B < 1.$$

The surface particle scattering is assumed to possess a high degree of specularity. Hence, the model of infinite cylinder quantum well is used. The transport mean free path of the particle is more or of the same order as the wire diameter. The spin of the particle is not taken into account.

Purpose: The main goal of the paper is to study how the longitudinal magnetic field applied to the elliptic wire affects the energy spectrum and caustic of electrons (holes) with isotropic mass.

The structure of the paper is the following. The second section is devoted to mathematical aspects for the calculation of the eigenfunction, eigenvalue, and corresponding caustic coordinate of the BS and RS. The third section examines the HCS and HOS. The numerical results for the all hole states in Bi wire are discussed in section 4. The outline of the obtained results is given in section five.

## 2. WHISPERING GALLERY MODES
### A. Boundary states

Let's obtain eigenvalues and eigenfunctions of Schrodinger equation in the elliptic domain when the uniform magnetic field is applied along z axis. In this case Schrodinger equation reads[6]

$$\frac{\partial^2 \Psi}{\partial \xi^2} + \frac{\partial^2 \Psi}{\partial \varphi^2} + Ed^2\left[ch^2(\xi) - \cos^2(\varphi)\right]\Psi - i\frac{c^2}{2L_B^2}\left[\sin(2\varphi)\frac{\partial \Psi}{\partial \xi} + sh(2\xi)\frac{\partial \Psi}{\partial \varphi}\right] - \left(\frac{c^2}{4L_B^2}\right)^2\left[sh^2(2\xi) + \sin^2(2\varphi)\right]\Psi = 0$$

(2.1)

The relation between the rectangular and elliptic system of coordinates is
$x = c \cdot ch(\xi)\cos(\varphi)$ and $y = c \cdot sh(\xi)\sin(\varphi)$, where $c = R\sqrt{m_1 - m_2}/(m_1 m_2)^{1/4}$ is half of the distance between the ellipse focuses, $0 \leq \xi \leq +\infty$, $0 \leq \varphi \leq 2\pi$. $L_B = \sqrt{\hbar/eB}$ - magnetic length, $d = (R/\hbar)\sqrt{2(m_1 - m_2)}$ is short notation.

The wire boundary is supposed to be impenetrable for a particle. The boundary condition is

$$\Psi(\bar{\xi}, \varphi) = 0,$$

(2.2)

where $\bar{\xi}$ is the coordinate of the ellipse boundary such that $th(\bar{\xi}) = \sqrt{m_2/m_1}$. In this section, we solve equation (2.1) for the functions localized near the ellipse boundary. These functions correspond to particles moving periodically along the ellipse boundary. These states are called as "whispering gallery" modes[5][16]. The second boundary condition is that the eigenfunction is single valued

$$\Psi(\xi, \varphi + 2\pi) = \Psi(\xi, \varphi). \tag{2.3}$$

Firstly, we expand functions $ch^2(\xi)$, $sh(2\xi)$, $sh^2(2\xi)$ in terms of variable $\xi$ at boundary $\bar{\xi}$ in equation (2.1). To solve equation (2.1) we use the following trial function[7]

$$\Psi(\xi, \varphi) = U(\nu, \varphi) = C \exp\left[i \sum_{m=-3}^{M} \alpha_m(\nu, \varphi) \omega^{-\frac{m}{3}}\right] \vartheta\left[\sum_{m=0}^{M} \beta_m(\nu, \varphi) \omega^{-\frac{m}{3}}\right], \tag{2.4}$$

where dimensionless eigenvalue $\omega = 2d\sqrt{E}$ is supposed to be large, $\alpha_m(\nu, \varphi)$ and $\beta_m(\nu, \varphi)$ are polynomials in $\nu$. $\nu = \omega^{2/3}(\xi - \bar{\xi})$ is new variable corresponding to the reduced distance from the point, $\bar{\xi}$, to the ellipse boundary, $\vartheta(Z)$ is the Airy function[17] ($\vartheta(Z) = \sqrt{\pi} Ai(Z)$). Formula (2.4) is substituted in equation (2.1). We use the following property of the Airy function $\vartheta''(Z) = Z\vartheta(Z)$. We rewrite equation (2.1) collecting the terms at $\vartheta(Z)$ and $\vartheta'(Z)$

$$a\left(\alpha_j, \beta_j; \omega^{-1/3}\right)\vartheta(Z) + b\left(\alpha_j, \beta_j; \omega^{-1/3}\right)\vartheta'(Z) = 0, \tag{2.5}$$

where $Z = \sum_{m=0}^{M} \beta_m(\nu, \varphi) \omega^{\frac{-m}{3}}$

$$a\left(\alpha_j, \beta_j; \omega^{-1/3}\right) = \omega^{4/3}\left[\sum_{m=-6}^{M-1} a_m(\alpha_j, \beta_j) \omega^{\frac{-m}{3}} + O\left(\omega^{\frac{-M}{3}}\right)\right],$$

$$b\left(\alpha_j, \beta_j; \omega^{-1/3}\right) = \omega^{4/3}\left[\sum_{m=-3}^{M-1} b_m(\alpha_j, \beta_j) \omega^{\frac{-m}{3}} + O\left(\omega^{\frac{-M}{3}}\right)\right]$$

Functions $\vartheta(Z)$ and $\vartheta'(Z)$ are linear independent. Therefore, coefficients $a\left(\alpha_j, \beta_j; \omega^{-1/3}\right)$ and $b\left(\alpha_j, \beta_j; \omega^{-1/3}\right)$ should be equal to zero concomitantly in equation (2.5). The following system of equations for polynomials $\alpha_j$ and $\beta_j$ is obtained

$$a_m(\alpha_j, \beta_j) = 0, \quad m \geq -6, \tag{2.6}$$
$$b_m(\alpha_j, \beta_j) = 0, \quad m \geq -3. \tag{2.7}$$

Letting $m$=-6, -5, -4 in equation (2.6) we obtain

$$a_{-6}(\alpha_j, \beta_j) = -\left(\frac{\partial \alpha_{-3}}{\partial \nu}\right)^2 = 0,$$

$$a_{-5}(\alpha_j, \beta_j) = -2\left(\frac{\partial \alpha_{-3}}{\partial \nu}\right)\left(\frac{\partial \alpha_{-2}}{\partial \nu}\right) = 0,$$

$$a_{-4}(\alpha_j, \beta_j) = -\left(\frac{\partial \alpha_{-2}}{\partial v}\right)^2 - 2\left(\frac{\partial \alpha_{-3}}{\partial v}\right)\left(\frac{\partial \alpha_{-1}}{\partial v}\right) = 0.$$

From these equations it follows that

$$\frac{\partial \alpha_{-3}}{\partial v} = \frac{\partial \alpha_{-2}}{\partial v} = 0, \tag{2.8}$$

i.e., $\alpha_{-2} = \alpha_{-2}(\varphi)$, $\alpha_{-3} = \alpha_{-3}(\varphi)$. Taking into account equation (2.8) we can write the following six equations from system (2.6) and (2.7) at $m = -3, -2, -1$ as

$$a_{-3}(\alpha_j, \beta_j) = 0,$$

$$a_{-2}(\alpha_j, \beta_j) = -\left(\frac{\partial \alpha_{-1}}{\partial v}\right)^2 - \left(\frac{\partial \alpha_{-3}}{\partial \varphi}\right) + \frac{1}{4}\left[ch^2(\bar{\xi}) - \cos^2(\varphi)\right] = 0,$$

$$a_{-1}(\alpha_j, \beta_j) = i\left(\frac{\partial^2 \alpha_{-1}}{\partial v^2}\right) - 2\left(\frac{\partial \alpha_{-3}}{\partial \varphi}\right)\left(\frac{\partial \alpha_{-2}}{\partial \varphi}\right) - 2\left(\frac{\partial \alpha_{-1}}{\partial v}\right)\left(\frac{\partial \alpha_{0}}{\partial v}\right) = 0,$$

$$b_{-3}(\alpha_j, \beta_j) = 0, \tag{2.9}$$

$$b_{-2}(\alpha_j, \beta_j) = 0,$$

$$b_{-1}(\alpha_j, \beta_j) = i2\left(\frac{\partial \alpha_{-1}}{\partial v}\right)\left(\frac{\partial \beta_0}{\partial v}\right) = 0.$$

The last equation from system (2.9) is satisfied when $\frac{\partial \alpha_{-1}}{\partial v} = 0$ or $\frac{\partial \beta_0}{\partial v} = 0$. Supposing that $\frac{\partial \beta_0}{\partial v} = 0$, we get to contradiction with the known solution of equation (2.1) for the circle boundary when the magnetic field is absent[7]. Hence, we let $\frac{\partial \alpha_{-1}}{\partial v} = 0$. Then we obtain from system (2.9)

$$\left(\frac{\partial \alpha_{-3}}{\partial \varphi}\right)^2 = \frac{1}{4}\left[ch^2(\bar{\xi}) - \cos^2(\varphi)\right], \tag{2.10}$$

$$\frac{\partial \alpha_{-2}}{\partial \varphi} = 0. \tag{2.11}$$

From equation (2.10) we obtain

$$\alpha_{-3}(v, \varphi) = \alpha_{-30}(\varphi) = \pm \frac{1}{2}\int_{d_{-3}}^{\varphi} \sqrt{ch^2(\bar{\xi}) - \cos^2(\tau)}\, d\tau, \tag{2.12}$$

where $d_{-3}$ is arbitrary constant. Two signs in (2.12) get two solutions of equation (2.1) corresponding to the motion of the particle in clockwise and anticlockwise manner along the boundary.

From relations (2.8) and (2.11) we obtain that $\alpha_{-2}$ is a constant

$$\alpha_{-2} = d_{-2}. \tag{2.13}$$

Taking into account relations (2.6-2.8) and (2.13) we write the next two equations of system (2.6) and (2.7) at $m = 0$

$$a_0(\alpha_j,\beta_j)=i\left(\frac{\partial^2\alpha_0}{\partial v^2}\right)-\left(\frac{\partial\alpha_0}{\partial v}\right)^2+\beta_0\left(\frac{\partial\beta_0}{\partial v}\right)^2-2\left(\frac{\partial\alpha_{-3}}{\partial\varphi}\right)\left(\frac{\partial\alpha_{-1}}{\partial\varphi}\right)+\frac{v}{4}sh(2\bar{\xi})=0, \quad (11.1)$$

$$b_0(\alpha_j,\beta_j)=i2\left(\frac{\partial\alpha_0}{\partial v}\right)\left(\frac{\partial\beta_0}{\partial v}\right)+\left(\frac{\partial^2\beta_0}{\partial v^2}\right)=0. \quad (2.14)$$

We eliminate derivatives $\frac{\partial\alpha_0}{\partial v}$ and $\frac{\partial^2\alpha_0}{\partial v^2}$ using (2.14). We get

$$-2\left(\frac{\partial^3\beta_0}{\partial v^3}\right)\left(\frac{\partial\beta_0}{\partial v}\right)+3\left(\frac{\partial^2\beta_0}{\partial v^2}\right)^2+4\beta_0\left(\frac{\partial\beta_0}{\partial v}\right)^4 \mp 4\left(\frac{d\alpha_{-1}}{d\varphi}\right)\left(\frac{\partial\beta_0}{\partial v}\right)^2\sqrt{ch^2(\bar{\xi})-\cos^2(\varphi)}+$$

$$v\cdot sh(2\bar{\xi})\left(\frac{\partial\beta_0}{\partial v}\right)^2=0$$

(2.15)

Let's suppose that $\beta_0(v,\varphi)$ is polynomial in $v$ of order $l$

$$\beta_0(v,\varphi)=\beta_{0l}(\varphi)v^l+...+\beta_{00}(\varphi). \quad (2.16)$$

The left-hand side of equation (2.15) is a polynomial of order $5l-4$ when $l\geq 2$

$$l^4\beta_{0l}^5(\varphi)v^{5l-4}+...=0. \quad (2.17)$$

Equation (2.17) is satisfied if all coefficients are equal to zero.

$$\beta_{0l}(\varphi)=0 \text{ if } l\geq 2.$$

Therefore, $\beta_0(v,\varphi)$ may be only polynomial of the first order in $v$, i.e.,

$$\beta_0(v,\varphi)=\beta_{01}(\varphi)v+\beta_{00}(\varphi). \quad (2.18)$$

Substituting expression (2.18) into equation (2.15), we obtain

$$\left[4\beta_{01}^3(\varphi)+sh(2\bar{\xi})\right]\cdot v+4\beta_{00}(\varphi)\beta_{01}^2(\varphi)\mp 4\left(\frac{d\alpha_{-1}}{d\varphi}\right)\sqrt{ch^2(\bar{\xi})-\cos^2(\varphi)}=0. \quad (2.19)$$

From the above equation, we obtain

$$\beta_{01}(\varphi)\equiv\beta_{01}=-\left[\frac{sh(2\bar{\xi})}{4}\right]^{1/3}.$$

Function $\beta_{00}(\varphi)$ will be obtained from the boundary conditions lately. The last term in equation (2.19) should be equal to zero.

$$\alpha_{-1}(v,\varphi)=\alpha_{-10}(\varphi)=\pm\beta_{01}\int_{d_{-1}}^{\varphi}\frac{\beta_{00}(\tau)d\tau}{\sqrt{ch^2(\bar{\xi})-\cos^2(\tau)}}. \quad (2.20)$$

From equation (2.14), it follows that $\alpha_0(v,\varphi)$ is a polynomial of the zero order because $\beta_0(v,\varphi)$ is the polynomial of the first order.

$$\alpha_0(v,\varphi)=\alpha_{00}(\varphi).$$

To find polynomials $\alpha_1(v,\varphi)$, $\beta_1(v,\varphi)$ and function $\alpha_{00}(\varphi)$ we consider the equations of system (2.6) and (2.7) at $m=1$. Taking into account relations (2.12), (2.13), (2.18), (2.20) we get

$$i\frac{\partial^2 \alpha_1}{\partial v^2} + 2\beta_{01}(\beta_{01}v + \beta_{00})\frac{\partial \beta_1}{\partial v} + \beta_{01}^2 \beta_1 \mp \sqrt{ch^2(\bar{\xi}) - \cos^2(\varphi)}\frac{d\alpha_{00}}{d\varphi} =$$
$$\mp \frac{i}{2}\frac{\cos(\varphi)\sin(\varphi)}{\sqrt{ch^2(\bar{\xi}) - \cos^2(\varphi)}} \mp \left(\frac{c}{2R_c}\right)^2 sh(2\bar{\xi})\sqrt{ch^2(\bar{\xi}) - \cos^2(\varphi)}$$
(2.21)

$$i2\beta_{01}\frac{\partial \alpha_1}{\partial v} + \frac{\partial^2 \beta_1}{\partial v^2} = \mp i\sqrt{ch^2(\bar{\xi}) - \cos^2(\varphi)}\frac{d\beta_{00}}{d\varphi}.$$
(2.22)

Eliminating function $\alpha_1(v,\varphi)$ from equation (2.21) we obtain

$$-\frac{\partial^3 \beta_1}{\partial v^3} + 4\beta_{01}^2(\beta_{01}v + \beta_{00})\frac{\partial \beta_1}{\partial v} + 2\beta_{01}^3 \beta_1 \mp 2\beta_{01}\sqrt{ch^2(\bar{\xi}) - \cos^2(\varphi)}\frac{d\alpha_{00}}{d\varphi} =$$
$$\mp i\beta_{01}\frac{\cos(\varphi)\sin(\varphi)}{\sqrt{ch^2(\bar{\xi}) - \cos^2(\varphi)}} \mp \left(\frac{c}{L_B}\right)^2 \frac{\beta_{01}}{2} sh(2\bar{\xi})\sqrt{ch^2(\bar{\xi}) - \cos^2(\varphi)}$$
(2.23)

Let's suppose that $\beta_1(v,\varphi)$ is polynomial of the $l$th order in form (2.16). After substituting into equation (2.23) we obtain that it can be only polynomial of the zero order.
$$\beta_1(v,\varphi) = \beta_{10}(\varphi). \tag{2.24}$$

Function $\beta_{10}(\varphi)$ is defined from the boundary conditions. Both sides of equation (2.23) depend only on variable $\varphi$ because $\beta_1$ does not depend on $v$. Integrating both sides of the equation, we obtain

$$\alpha_{00}(\varphi) = \pm \beta_{01}\int_{d_0}^{\varphi} \frac{\beta_{10}(\tau)d\tau}{\sqrt{ch^2(\bar{\xi}) - \cos^2(\tau)}} + \frac{i}{2}\ln\left[\sqrt{ch^2(\bar{\xi}) - \cos^2(\varphi)}\right] + \left(\frac{c}{2L_B}\right)^2 sh(2\bar{\xi})(\varphi - d_0),$$
(2.25)

where $d_0$ is arbitrary constant.

Substituting (2.24) into (2.22) and integrating this equation with respect to variable $v$ we find $\alpha_1(v,\varphi)$ which is polynomial of the first order

$$\alpha_1(v,\varphi) = \alpha_{11}(\varphi)v + \alpha_{10}(\varphi) = \mp \frac{\sqrt{ch^2(\bar{\xi}) - \cos^2(\varphi)}}{2\beta_{01}}\frac{d\beta_{00}}{d\varphi}v + \alpha_{10}(\varphi). \tag{2.26}$$

Polynomials $\alpha_2(v,\varphi)$, $\beta_2(v,\varphi)$ and function $\alpha_{10}(\varphi)$ are obtained from the system of equations (2.6) and (2.7) when $m=2$
$$\beta_2(v,\varphi) = \beta_{22}(\varphi)v^2 + \beta_{21}(\varphi)v + \beta_{20}(\varphi), \tag{2.27}$$

where
$$\beta_{22}(\varphi) \equiv \beta_{22} = -\frac{ch(2\bar{\xi})}{20}\left[\frac{4}{sh(2\bar{\xi})}\right]^{\frac{2}{3}},$$

$$\beta_{21}(\varphi) = -\frac{4}{3}\frac{\beta_{22}}{\beta_{01}}\beta_{00}(\varphi) \mp \frac{\cos(\varphi)\sin(\varphi)}{6\beta_{01}^3}\frac{d\beta_{00}}{d\varphi} \mp \frac{[ch^2(\bar{\xi}) - \cos^2(\varphi)]}{6\beta_{01}^3}\frac{d^2\beta_{00}}{d\varphi^2}.$$

$\beta_{20}(\varphi)$ is defined from the boundary conditions.

$$\alpha_{10}(\varphi) = \mp \frac{8}{3}\beta_{22}\int_{d_1}^{\varphi}\frac{\beta_{00}^2(\tau)d\tau}{\sqrt{ch^2(\bar{\xi})-\cos^2(\tau)}} - \frac{1}{3\beta_{01}^3}\int_{d_1}^{\varphi}\frac{\sin(\tau)\cos(\tau)\beta_{00}(\tau)}{\sqrt{ch^2(\bar{\xi})-\cos^2(\tau)}}\left(\frac{d\beta_{00}}{d\tau}\right)d\tau -$$

$$\frac{1}{3\beta_{01}^2}\int_{d_1}^{\varphi}\beta_{00}(\tau)\sqrt{ch^2(\bar{\xi})-\cos^2(\tau)}\left(\frac{d^2\beta_{00}}{d\tau^2}\right)d\tau \pm \beta_{01}^2\int_{d_1}^{\varphi}\frac{\beta_{20}(\tau)d\tau}{\sqrt{ch^2(\bar{\xi})-\cos^2(\tau)}} \mp \qquad (2.28)$$

$$\frac{1}{4\beta_{01}^2}\int_{d_1}^{\varphi}\sqrt{ch^2(\bar{\xi})-\cos^2(\tau)}\left(\frac{d\beta_{00}}{d\tau}\right)^2 d\tau \mp \int_{d_1}^{\varphi}\frac{\beta_{00}^2(\tau)d\tau}{[ch^2(\bar{\xi})-\cos^2(\tau)]^{3/2}}$$

Polynomial $\alpha_2(\nu,\varphi)$ is

$$\alpha_2(\nu,\varphi) = i\frac{\beta_{22}}{\beta_{01}}\nu \pm i\frac{\sqrt{ch^2(\bar{\xi})-\cos^2(\varphi)}}{2\beta_{01}}\frac{d\beta_{10}}{d\varphi}\nu + \left(\frac{c}{2L_B}\right)^2 \sin(2\varphi)\nu + \alpha_{20}(\varphi).$$

Function $\alpha_{20}(\varphi)$ and polynomials $\alpha_3(\nu,\varphi)$, $\beta_3(\nu,\varphi)$ are defined from system (2.6) and (2.7) when $m=3$.

The solution of equation (2.1) can be written as

$$U_{\pm}(\nu,\varphi) = const \cdot \exp\left\{i\left[\alpha_{-30}\omega + \alpha_{-2}\omega^{\frac{2}{3}} + \alpha_{-10}\omega^{\frac{1}{3}} + \alpha_{00} + \alpha_1\omega^{-\frac{1}{3}}\right]\right\} \vartheta\left(\beta_0 + \beta_1\omega^{-\frac{1}{3}} + \beta_2\omega^{-\frac{2}{3}}\right)$$

(2.29)

where polynomials
$\alpha_{-30}(\varphi)$, $\alpha_{-2}$, $\beta_0(\nu,\varphi)$, $\alpha_{-10}(\varphi)$, $\beta_1(\nu,\varphi)$, $\alpha_{00}(\varphi)$, $\alpha_1(\nu,\varphi)$, and $\beta_2(\nu,\varphi)$ are defined by means of formulas (2.12), (2.13), (2.18), (2.20), (2.24), (2.25), (2.26), (2.27). Substituting solution (2.29) into boundary condition (2.2) and reducing the exponent function, we obtain equation

$$\vartheta\left[\beta_0(\bar{\nu},\varphi) + \beta_{10}(\varphi)\omega^{-\frac{1}{3}} + \beta_2(\bar{\nu},\varphi)\omega^{-\frac{2}{3}}\right] = 0. \qquad (2.30)$$

Functions $\beta_{m0}(\varphi)$ are defined from equation (2.30)
$$\beta_{00}(\varphi) = -t_p \text{ when } p=1,2,\ldots$$
$$\beta_{m0}(\varphi) = 0 \text{ when } m \geq 1 \qquad (2.31)$$

The zeros of the Airy function[17] are $t_1=2.33811$, $t_2=4.08795$, $t_3=5.52056$, $t_4=6.78671$, $t_5=7.94417$. When the value of parameter $p$ is large, the following asymptotic formula can be applied

$$t_p = \left[\frac{3\pi}{2}\left(p - \frac{1}{4}\right)\right]^{2/3}$$

Terms $\alpha_{m0}(\varphi)$ of polynomial $\alpha_m(\nu,\varphi)$, when $m \geq -3$, are expressed by means of integrals with variable upper limit and constant lower limit. Choosing appropriate constant factor in solution (2.29), we can take all lower limits of integration $d_m$ being equal to zero i.e.,

$$\alpha_m(0,0)=0, \text{ when } m \geq -3 \qquad (2.32)$$

Taking into account conditions (2.31), (2.32), we obtain the following expressions for $\alpha_{m0}(\varphi)$ using formulas (2.12), (2.13), (2.20), (2.25), (2.28) when $m=-3, -2, -1, 0, 1$

$$\alpha_{-30}(\varphi) = \pm \int_0^\varphi \sqrt{ch^2(\bar{\xi}) - \cos^2(\tau)}\, d\tau,$$

$$\alpha_{-20} = 0,$$

$$\alpha_{-10}(\varphi) = \mp t_p \left[\frac{sh(2\bar{\xi})}{4}\right]^{2/3} \int_0^\varphi \frac{d\tau}{\sqrt{ch^2(\bar{\xi}) - \cos^2(\tau)}},$$

$$\alpha_{00}(\varphi) = \frac{i}{4}\ln[ch^2(\bar{\xi}) - \cos^2(\varphi)] + \left(\frac{c}{2L_B}\right)^2 sh(2\bar{\xi})\varphi,$$

$$\alpha_{10}(\varphi) = \mp t_p^2 \left[\frac{sh(2\bar{\xi})}{4}\right]^{4/3} \int_0^\varphi \frac{d\tau}{[ch^2(\bar{\xi}) - \cos^2(\tau)]^{3/2}} \pm \frac{2}{15} t_p^2 ch(2\bar{\xi}) \left[\frac{4}{sh(2\bar{\xi})}\right]^{2/3} \int_0^\varphi \frac{d\tau}{\sqrt{ch^2(\bar{\xi}) - \cos^2(\tau)}}$$

One can show that $c^2 sh(2\bar{\xi}) = 2R^2$. The solution of equation (2.1) is written in the second order approximation as

$$U_p^\pm(\nu, \varphi) = \frac{const}{\sqrt[4]{ch^2(\bar{\xi}) - \cos^2(\varphi)}} \exp\left\{\pm i \frac{\omega}{2} \int_0^\varphi \sqrt{ch^2(\bar{\xi}) - \cos^2(\tau)}\, d\tau\right\} \times$$

$$\exp\left\{\mp i\omega^{\frac{1}{3}} t_p \left(\frac{R}{c\sqrt{2}}\right)^{4/3} \int_0^\varphi \frac{d\tau}{\sqrt{ch^2(\bar{\xi}) - \cos^2(\tau)}} + i\left(\frac{R}{L_B\sqrt{2}}\right)^2 \varphi \mp i\omega^{\frac{-1}{3}} t_p^2 \left(\frac{R}{c\sqrt{2}}\right)^{8/3} \int_0^\varphi \frac{d\tau}{[ch^2(\bar{\xi}) - \cos^2(\tau)]^{3/2}}\right\} \times$$

$$\exp\left\{\pm i\omega^{\frac{-1}{3}} \frac{2t_p^2 ch(2\bar{\xi})}{15}\left(\frac{c\sqrt{2}}{R}\right)^{4/3} \int_0^\varphi \frac{d\tau}{\sqrt{ch^2(\bar{\xi}) - \cos^2(\tau)}}\right\} \times$$

$$\vartheta\left\{-t_p - \left(\frac{R}{c\sqrt{2}}\right)^{2/3}\nu - \omega^{\frac{-2}{3}}\left(\frac{c\sqrt{2}}{R}\right)^{4/3} \frac{ch(2\bar{\xi})}{20}\nu^2 + \omega^{\frac{-2}{3}} \frac{4}{15} t_p cth(2\bar{\xi})\nu\right\}. \qquad (2.33)$$

If function $f(x)$ is even, periodic function with period $2\pi$, i.e., $f(-x) = f(x)$ and $f(x + 2\pi) = f(x)$, then

$$\int_0^{\varphi + 2\pi} f(x)\, dx = \int_0^{2\pi} f(x)\, dx + \int_0^\varphi f(x)\, dx. \qquad (2.34)$$

Solution (2.33) is substituted into the second boundary condition given by formula (2.3). Taking into consideration property (2.34), we obtain the second quantization condition for the energy spectrum

$$\frac{\omega_{p,q}}{2} \int_0^{2\pi} \sqrt{ch^2(\bar{\xi}) - \cos^2(\tau)}\, d\tau - \omega_{p,q}^{1/3}\left(\frac{R}{c\sqrt{2}}\right)^{4/3} t_p \int_0^{2\pi} \frac{d\tau}{\sqrt{ch^2(\bar{\xi}) - \cos^2(\tau)}} \pm \left(\frac{R}{L_B}\right)^2 \pi -$$

$$\omega_{p,q}^{-1/3}\left(\frac{R}{c\sqrt{2}}\right)^{8/3} t_p^2 \int_0^{2\pi} \frac{d\tau}{[ch^2(\bar{\xi}) - \cos^2(\tau)]^{3/2}} + \omega_{p,q}^{-1/3} \frac{2}{15}\left(\frac{c\sqrt{2}}{R}\right)^{4/3} t_p^2 ch(2\bar{\xi}) \int_0^{2\pi} \frac{d\tau}{\sqrt{ch^2(\bar{\xi}) - \cos^2(\tau)}} = 2\pi q$$

(2.35)

where $q$ is an integer corresponding to angular momentum quantum number such that $q \gg p$. Formula (2.35) can be reduced to the result obtained by Boghachek[16] for the energy spectrum of a particle moving in the circular cylinder under applied longitudinal magnetic field.

The Airy function $\vartheta(Z)$ oscillates when Z<0 and exponentially decreases when Z>0. Hence, function $U_{p,q}(v,\varphi)$ oscillates in the ring defined by
$$v_{ec} < v \leq 0$$
Elliptic caustic coordinate $v_{ec}$ is defined from the equation
$$-t_p + \beta_{01} v_{ec} + \omega_{p,q}^{-\frac{2}{3}} \beta_{22} v_{ec}^2 + \omega_{p,q}^{-\frac{2}{3}} \left( \frac{4\beta_{22}}{3\beta_{01}} \right) t_p v_{ec} = 0. \quad (2.36)$$

The solution of equation (2.36) in the first order of approximation is
$$v_{ec} = -t_p \left( \frac{c\sqrt{2}}{R} \right)^{\frac{2}{3}}$$

Finally
$$\xi_{ec} = \overline{\xi} - t_p \left( \frac{\hbar^2}{4\sqrt{m_1 m_2} E_{p,q} R^2} \right)^{\frac{1}{3}}. \quad (2.37)$$

Formula (2.37) shows that the large angular momentum quantum number $q$ causes the narrowing of the boundary elliptic ring accessible for electron (hole) motion.

Therefore, for the particle moving along the elliptic boundary, expression (2.33) and (2.35) define the eigenfunction and eigenvalue of equation (2.1). The influence of the magnetic field on the elliptic caustic coordinate is estimated by means of eigenvalue $\omega_{p,q}$ from formula (2.36). The effect of the magnetic field on the elliptic caustic is uniform along whole length of the caustic.

**B. Ring states**

Let's consider the states of a particle moving in the vicinity of the elliptic caustic. To solve Schrodinger equation (2.1), we expand functions $ch^2(\xi)$, $sh(2\xi)$, $sh^2(2\xi)$ in terms of $\xi$ at elliptic caustic coordinate $\xi_{ec}$. Solution of equation (2.1) is given by trial function (2.4) where $v = \omega^{2/3} (\xi - \xi_{ec})$. Applying the method described in the previous subsection, we get the similar results where parameter $\overline{\xi}$ is replaced by $\xi_{ec}$.

Solution (2.29) contains the Airy function $\vartheta(Z)$. Argument Z of the Airy function is equal to zero at the caustic. Hence, terms $\beta_{m0}(\varphi)$, where $m \geq 0$, are defined by equations
$$\beta_{m0}(\varphi)=0, \text{ if } m \geq 0 \quad (2.38)$$
The first quantization condition for energy spectrum is obtained from boundary condition (2.2) using relations (2.38)

$$\omega_{p,q}^{2/3}\left[\frac{sh(2\xi_{ec})}{4}\right]^{\frac{1}{3}}\left[(\bar{\xi}-\xi_{ec})+\frac{cth(2\xi_{ec})}{5}(\bar{\xi}-\xi_{ec})^2\right]=t_p. \qquad (2.39)$$

One can demonstrate that condition (2.39) presents the series expansion in terms of $(\bar{\xi}-\xi_{ec})$, limited to its first two terms, for more general condition

$$\frac{3}{4}\omega_{p,q}\int_{\xi_{ec}}^{\bar{\xi}}\sqrt{ch^2(\xi)-ch^2(\xi_{ec})}d\xi=t_p^{3/2} \qquad (2.40)$$

In the second order approximation, the solution of equation (2.1) is

$$U^{\pm}(v,\varphi)=\frac{const}{\sqrt[4]{ch^2(\xi_{ec})-\cos^2(\varphi)}}\exp\left\{\pm i\frac{\omega}{2}\int_0^{\varphi}\sqrt{ch^2(\xi_{ec})-\cos^2(\tau)}d\tau+i\left(\frac{c}{2L_B}\right)^2 sh(2\xi_{ec})\varphi\right\}\times$$

$$\vartheta\left\{-\left(\frac{sh(2\xi_{ec})}{4}\right)^{1/3}v-\omega^{-2/3}\left(\frac{4}{sh(2\xi_{ec})}\right)^{2/3}\frac{ch(2\xi_{ec})}{20}v^2\right\}$$

(2.41)

Using function (2.41), we obtain the second quantization condition for the energy spectrum from relation (2.3)

$$\frac{\omega_{p,q}}{2}\int_0^{2\pi}\sqrt{ch^2(\xi_{ec})-\cos^2(\tau)}d\tau\pm\left(\frac{c}{L_B}\right)^2\frac{\pi\cdot sh(2\xi_{ec})}{2}=2\pi q, \qquad (2.42)$$

where $q$ is integer angular momentum quantum number such that $q>>p$. The first integral in formula (2.42) corresponds to the elliptic caustic length. The second term gives the ratio of the area bounded by the elliptic caustic, $S_e=\pi\cdot c^2 sh(\xi_{ec})ch(\xi_{ec})$, to the magnetic length $L_B$. Nedorezov[6] obtained the quantization conditions for energy spectrum similar with expressions (2.40), (2.42) in the framework of perturbation theory. Eigenvalue, $\omega_{p,q}$, of equation (2.1) and elliptic caustic coordinate $\xi_{ec}$ are calculated from the system of nonlinear equations (2.40) and (2.42). While in the case of boundary state, the eigenvalue and caustic coordinate are obtained from two distinct equations. Eigenfunction (2.40) and eigenvalue $\omega_{p,q}$ of equation (2.1) correspond to a particle moving along rays in the elliptic ring between the boundary $\bar{\xi}$ and elliptic caustic $\xi_{ec}$.

## 3. JUMPING BALL MODES
### A. Hyperbolic caustic states

Let's study motion of the particle with small angular momentum quantum number and large radial quantum number in the elliptic wire cross section under applied weak longitudinal magnetic field. Our scope in this section is to find an asymptotic solution of the Schrodinger equation in the wire cross section core. In the absence of the magnetic field, this solution corresponds to particles moving between two focuses of the boundary ellipse and reflecting under a big tilt angle toward the boundary. The hyperbolic caustics and elliptic boundary restrict the motion of the particles[5]. The corresponding states of the particle are attributed to "jumping ball" modes introduced in the short-wave length diffraction theory[7]. In this subsection, we pay attention to the solution defined in the vicinity of the hyperbolic caustic. Further, for convenience, we use a new elliptic system of coordinates

$$x = c \cdot ch(\xi)\sin(\varphi), \quad y = c \cdot sh(\xi)\cos(\varphi),$$

where $-\infty < \xi < +\infty$ and $-\pi/2 < \varphi < \pi/2$. Schrodinger equation reads in this coordinate system as

$$\frac{\partial^2 \Psi}{\partial \xi^2} + \frac{\partial^2 \Psi}{\partial \varphi^2} + Ed^2\left[ch^2(\xi) - \sin^2(\varphi)\right]\Psi - i\frac{c^2}{2L_B^2}\left[\sin(2\varphi)\frac{\partial \Psi}{\partial \xi} - sh(2\xi)\frac{\partial \Psi}{\partial \varphi}\right] - \left(\frac{c^2}{4L_B^2}\right)^2\left[sh^2(2\xi) + \sin^2(2\varphi)\right]\Psi = 0 \quad (3.1)$$

The boundary condition is

$$\Psi(\pm \bar{\xi}, \varphi) = 0. \quad (3.2)$$

The operation of complex conjugation and simultaneous replacement $\varphi \to -\varphi$ do not change equation (3.1). Therefore, the following property for the eigenfunction is stated

$$\Psi(\xi, \varphi) = \pm \Psi^*(\xi, -\varphi). \quad (3.3)$$

From formula (3.3) we can deduce the second boundary condition

$$\begin{aligned}\text{Re}\{\Psi(\xi, 0)\} &= 0 \\ \text{Re}\{\Psi'_\varphi(\xi, 0)\} &= 0\end{aligned} \quad (3.4)$$

At hyperbolic caustic coordinate $\varphi_{hc}$, functions $\sin^2(\varphi)$, $\sin(2\varphi)$, $\sin^2(2\varphi)$ are expanded in terms of $\varphi$. To solve equation (3.1) we use trial function (2.4). A new variable $v$ is defined as $v = \omega^{2/3}(\varphi - \varphi_{hc})$. Using the method described in section 2.B we obtain the solution of equation (3.1) in the form

$$U_\pm(\xi, v) = \frac{const}{\sqrt[4]{ch^2(\xi) - \sin^2(\varphi_{hc})}} \exp\left\{\pm i\frac{\omega}{2}\int_0^\xi \sqrt{ch^2(\tau) - \sin^2(\varphi_{hc})}d\tau + i\left(\frac{c}{2L_B}\right)^2 \sin(2\varphi_{hc})\xi\right\} \times$$

$$\vartheta\left\{\left(\frac{\sin(2\varphi_{hc})}{4}\right)^{1/3} v + \omega^{-2/3}\left(\frac{4}{\sin(2\varphi_{hc})}\right)^{2/3} \frac{\cos(2\varphi_{hc})}{20}v^2\right\}$$

(3.5)

The first quantization condition for energy spectrum is obtained substituting solution (3.5) into boundary condition (3.4)

$$\omega_{p,q}^{2/3}\left[\frac{\sin(2\varphi_{hc})}{4}\right]^{1/3}\varphi_{hc}\left[1 - \frac{ctg(2\varphi_{hc})}{5}\varphi_{hc}\right] = \begin{Bmatrix} t_q \\ t'_q \end{Bmatrix}, \quad (3.6)$$

where $q = 1, 2, 3, \ldots$  $t'_q$ is root of the equation $\vartheta'(Z) = 0$. The first five values of parameter $t'_q$ are[17][18] $t'_1 = 1.01879$, $t'_2 = 3.24820$, $t'_3 = 4.82010$, $t'_4 = 6.16331$, $t'_5 = 7.37218$. At large value of index $p$, the asymptotic formula is $t'_q = [(3\pi/2)(q - 3/4)]^{2/3}$. The 1st equation from system (3.6) corresponds to the even hyperbolic caustic states (HCS), the 2nd to the odd HCS.

The general solution of equation (3.1) is

$$\Psi(\xi,\varphi) = A_+ \Psi_+(\xi,\varphi) + A_- \Psi_-(\xi,\varphi). \tag{3.7}$$

Substituting solution (3.7) into boundary condition (3.2), we obtain the following system of linear algebraic equations for unknowns $A_-$ and $A_+$

$$A_+ \exp\left[i\frac{\omega}{2}\int_0^{\bar{\xi}} \sqrt{ch^2(\tau) - \sin^2(\varphi_{hc})}d\tau\right] + A_- \exp\left[-i\frac{\omega}{2}\int_0^{\bar{\xi}} \sqrt{ch^2(\tau) - \sin^2(\varphi_{hc})}d\tau\right] = 0$$

$$A_+ \exp\left[-i\frac{\omega}{2}\int_0^{\bar{\xi}} \sqrt{ch^2(\tau) - \sin^2(\varphi_{hc})}d\tau\right] + A_- \exp\left[i\frac{\omega}{2}\int_0^{\bar{\xi}} \sqrt{ch^2(\tau) - \sin^2(\varphi_{hc})}d\tau\right] = 0$$

$$\tag{3.8}$$

Given system of equations has nontrivial solution if the determinant of the coefficients of system (3.8) is equal to zero

$$i2\sin\left(\omega\int_0^{\bar{\xi}} \sqrt{ch^2(\tau) - \sin^2(\varphi_{hc})}d\tau\right) = 0$$

So, we obtain the second quantization condition for the energy spectrum

$$\omega_{p,q} \int_0^{\bar{\xi}} \sqrt{ch^2(\tau) - \sin^2(\varphi_{hc})}d\tau = \pi p, \tag{3.9}$$

where $p=1,2,3,\ldots$ is an integer such that $p \gg q$. For quantum resonators designed on the base of elliptic mirrors in the magnetic field absence, formulas (3.6) and (3.9) can be reduced to the quantization condition obtained by Bykov et al.[9]

Using expressions (3.8) and (3.9) we get the following relation between coefficients $A_-$ and $A_+$

$$A_+ = (-1)^{p+1} A_- \tag{3.10}$$

Using formula (3.10) we write the general solution of equation (3.1) as follows

$$\Psi_{p,q}(\xi,\varphi) = \frac{const}{\sqrt[4]{ch^2(\xi) - \sin^2(\varphi_{hc})}} \exp\left[i\left(\frac{c}{2L_B}\right)^2 \sin(2\varphi_{hc})\xi\right] \begin{Bmatrix} \sin \\ \cos \end{Bmatrix} \left\{\frac{\omega_{p,q}}{2}\int_0^{\bar{\xi}} \sqrt{ch^2(\tau) - \sin^2(\varphi_{hc})}d\tau\right\} \times$$

$$\vartheta\left\{\omega_{p,q}^{2/3}\left[\frac{\sin(2\varphi_{hc})}{4}\right]^{\frac{1}{3}}\left[(\varphi - \varphi_{hc}) + \frac{ctg(2\varphi_{hc})}{5}(\varphi - \varphi_{hc})^2\right]\right\}$$

$$\tag{3.11}$$

In formula (3.11), sin (cos) is taken if $p$ is even (odd). Large number $p$ is the number of a half wave oscillation of function (3.11) along axis $Y$ and $q$ is number of its half wave oscillations along axis $X$. The eigenvalue obtained from the first equation of system (3.6) corresponds to an even number of the eigenfunction oscillations along $X$ axis. The second equation corresponds to an odd number of the eigenfunction oscillations.

Therefore, the HCS eigenvalue of equation (3.1) and value of the corresponding hyperbolic caustic coordinate are obtained from the system of nonlinear equations (3.6) and (3.9). The HCS eigenfunction is given by formula (3.11). In the first order approximation, for small value of the angular momentum quantum number, the HCS energy of the particle moving between two hyperbolic caustics as well as the hyperbolic caustic coordinate itself are independent of magnetic field.

## B. Harmonic oscillator states

We consider a particle having a little magnetic quantum number. The HOS particle is supposed to be more localized at the least ellipse axis in comparison with that corresponding to the hyperbolic caustic state. In this case, if magnetic field is neglected, then equation (2.1) is reduced to the equation of the harmonic oscillator [12].

We rewrite equation (3.1) using new variable $v = \omega^{1/2} \sin(\varphi)$

$$\frac{\partial^2 \Psi}{\partial \xi^2} + \omega\left(1 - \frac{v^2}{\omega}\right)\frac{\partial^2 \Psi}{\partial v^2} - v\frac{\partial \Psi}{\partial v} + \frac{\omega^2}{4}\left[ch^2(\xi) - \frac{v^2}{\omega}\right]\Psi - i\left(\frac{c}{\sqrt{2}L_B}\right)^2 \sqrt{1 - \frac{v^2}{\omega}} \times $$
$$\left[\frac{2v}{\sqrt{\omega}}\frac{\partial \Psi}{\partial \xi} - sh(2\xi)\sqrt{\omega}\frac{\partial \Psi}{\partial v}\right] - \left(\frac{c}{2L_B}\right)^4\left[sh^2(2\xi) + \frac{4v^2}{\omega}\left(1 - \frac{v^2}{\omega}\right)\right]\Psi = 0 \quad (3.12)$$

One expands function $\sqrt{1 - v^2/\omega}$ in terms of $v^2/\omega$ at zero point. To solve equation (3.12) we use the trial function[7]

$$\Psi(\xi, \varphi) = U(\xi, v) = \exp\left[i\sum_{m=-2}^{M-1}\alpha_m(\xi, v)\omega^{\frac{-m}{2}}\right]D_q\left[\sqrt{2}\sum_{m=0}^{M-1}\beta_m(\xi, v)\omega^{\frac{-m}{2}}\right], \quad (3.13)$$

where $D_q(\sqrt{2}Z)$ is function of parabolic cylinder[19] which satisfies the following equation

$$\frac{d^2 D_q(\sqrt{2}Z)}{dZ^2} + [(2q+1) - Z^2]D_q(\sqrt{2}Z) = 0.$$

Functions $\alpha_m(\xi, v)$, $\beta_m(\xi, v)$ are polynomials of order $m$ in $v$. Substituting trial function (3.13) into equation (3.12), we obtain the following expression

$$a(\xi, v; \omega)D_q\left[\sqrt{2}Z(\xi, v, \omega)\right] + b(\xi, v; \omega)\frac{dD_q\left[\sqrt{2}Z(\xi, v, \omega)\right]}{dZ} = 0. \quad (3.14)$$

Equation (3.14) is satisfied if coefficients $a(\xi, v; \omega)$ and $b(\xi, v; \omega)$ at linear independent functions $D_q(\sqrt{2}Z)$ and $\frac{dD_q(\sqrt{2}Z)}{dZ}$ are equal to zero. Expanding functions $a(\xi, v; \omega)$ and $b(\xi, v; \omega)$ in series in terms of $\omega^{-1/2}$ we get

$$a(\xi,v;\omega) = \omega \sum_{m=-4}^{M-1} a_m(\xi,v)\omega^{\frac{-m}{2}} = 0,$$

$$b(\xi,v;\omega) = \omega \sum_{m=-2}^{M-1} b_m(\xi,v)\omega^{\frac{-m}{2}} = 0. \tag{3.15}$$

Coefficients $a_m(\xi,v)$ and $b_m(\xi,v)$ contain partial derivatives of polynomials $\alpha_m(\xi,v)$, $\beta_m(\xi,v)$. Equations (3.15) are satisfied if

$$a_m(\xi,v)=0, \quad m=-4, -3, -2, \ldots \tag{3.16}$$

$$b_m(\xi,v)=0, \quad m=-2, -1, 0, \ldots \tag{3.17}$$

Expressions (3.16) and (3.17) present the system of recurrent equations in partial derivatives of polynomials $\alpha_m(\xi,v)$, $\beta_m(\xi,v)$. Taking $m=-4,-3$ and $m=-2,-1$ in equations (3.16) and (3.17), correspondingly, we obtain

$$a_{-4}(\xi,v) = -\left(\frac{\partial \alpha_{-2}}{\partial v}\right)^2 = 0$$

$$a_{-3}(\xi,v) = -2\left(\frac{\partial \alpha_{-2}}{\partial v}\right)\left(\frac{\partial \alpha_{-1}}{\partial v}\right) = 0$$

$$b_{-2}(\xi,v) = i\sqrt{2}\left(\frac{\partial \alpha_{-2}}{\partial v}\right)\left(\frac{\partial \beta_0}{\partial v}\right) = 0$$

$$b_{-1}(\xi,v) = i\sqrt{2}\left(\frac{\partial \alpha_{-2}}{\partial v}\right)\left(\frac{\partial \beta_1}{\partial v}\right) - i\sqrt{2}v^2\frac{\partial \alpha_{-1}}{\partial v} = 0$$

The above equations show that $\alpha_{-2}(\xi,v)$ and $\alpha_{-1}(\xi,v)$ do not depend on $v$. Hence, they are polynomials of zero order in $v$.

$$\alpha_{-2}(\xi,v)=\alpha_{-20}(\xi),$$
$$\alpha_{-1}(\xi,v)=\alpha_{-10}(\xi). \tag{3.18}$$

Taking into account formulas (3.18), we write the following relations at $m=-2, -1$ from equation (3.16)

$$a_{-2}(\xi,v) = -\left(\frac{\partial \alpha_{-2}}{\partial \xi}\right)^2 + \frac{ch^2(\xi)}{4} = 0,$$

$$a_{-1}(\xi,v) = 2\left(\frac{\partial \alpha_{-2}}{\partial \xi}\right)\left(\frac{\partial \alpha_{-1}}{\partial \xi}\right) = 0.$$

From these equations it follows that

$$\alpha_{-2}(\xi,v) \equiv \alpha_{-20}(\xi) = \pm\frac{1}{2}[sh(\xi) - sh(d_{-2})], \tag{3.19}$$

$$\alpha_{-10}(\xi)=d_{-1},$$

where $d_{-1}$, $d_{-2}$ are arbitrary constants. Taking into consideration the obtained relations, we write the following set of equations from system (3.16) and (3.17) at $m=0$

$$a_0(\xi,v)=\pm i\frac{sh(\xi)}{2}\mp ch(\xi)\frac{\partial \alpha_0}{\partial \xi}+i\frac{\partial^2 \alpha_0}{\partial v^2}-\left(\frac{\partial \alpha_0}{\partial v}\right)^2+\left(\frac{\partial \beta_0}{\partial v}\right)^2\beta_0^2-(2q+1)\left(\frac{\partial \beta_0}{\partial v}\right)^2-\frac{v^2}{4}=0 \tag{3.20}$$

$$b_0(\xi,v)=i2\left(\frac{\partial \alpha_0}{\partial v}\right)\left(\frac{\partial \beta_0}{\partial v}\right)+\frac{\partial^2 \beta_0}{\partial v^2}\pm ich(\xi)\frac{\partial \beta_0}{\partial \xi}=0. \tag{3.21}$$

From equation (3.21), it follows that $\alpha_0(\xi,v)$ is polynomial of the second order in $v$

$$\alpha_0(\xi,v)=\alpha_{00}(\xi)+\alpha_{01}(\xi)v+\alpha_{02}(\xi)v^2. \tag{3.22}$$

From expressions (3.22) and (3.20), it follows that $\beta_0(\xi,v)$ is polynomial of the first order in $v$

$$\beta_0(\xi,v)=\beta_{00}(\xi)+\beta_{01}(\xi)v. \tag{3.23}$$

Substituting polynomials (3.22) and (3.23) into equations (3.20) and (3.21) we obtain

$$\left[\mp ch(\xi)\alpha'_{02}-4\alpha_{02}^2+\beta_{01}^4-1/4\right]\cdot v^2+\left[\mp ch(\xi)\alpha'_{01}-4\alpha_{01}\alpha_{02}+2\beta_{00}\beta_{01}^3\right]\cdot v \\ \pm i[sh(\xi)/2]\mp ch(\xi)\alpha'_{00}+i2\alpha_{02}-\alpha_{01}^2+\beta_{00}^2\beta_{01}^2-(2q+1)\beta_{01}^2=0 \tag{3.24}$$

$$[4\alpha_{02}\beta_{01}\pm\beta'_{01}ch(\xi)]\cdot v+2\alpha_{01}\beta_{01}\pm ch(\xi)\beta'_{00}=0. \tag{3.25}$$

Equation (3.25) is satisfied if the coefficients in front of $v$ and $v^0$ are equal to zero. Hence, we obtain

$$\alpha_{01}=\mp ch(\xi)\frac{\beta'_{00}}{2\beta_{01}},$$

$$\alpha_{02}=\mp ch(\xi)\frac{\beta'_{01}}{4\beta_{01}}. \tag{3.26}$$

Expressions (3.26) are substituted into (3.24). Taking into account that the coefficients in front of $v^2$, $v$ and $v^0$ are equal to zero, we obtain the equations for calculating $\beta_{00}$, $\beta_{01}$ and $\alpha_{00}$.

$$ch(\xi)\left[ch(\xi)\frac{\beta'_{01}}{4\beta_{01}}\right]'-4\left[ch(\xi)\frac{\beta'_{01}}{4\beta_{01}}\right]^2+\beta_{01}^4-\frac{1}{4}=0, \tag{3.27}$$

$$ch(\xi)\left[ch(\xi)\frac{\beta'_{00}}{2\beta_{01}}\right]'-4\alpha_{01}\alpha_{02}+2\beta_{00}\beta_{01}^3=0, \tag{3.28}$$

$$ch(\xi)\alpha'_{00}=i\frac{sh(\xi)}{2}+i2\alpha_{02}\mp ch^2(\xi)\left(\frac{\beta'_{00}}{2\beta_{01}}\right)^2\pm\beta_{01}^2\beta_{00}^2\mp(2q+1)\beta_{01}^2. \tag{3.29}$$

To solve equation (3.27) we introduce new function $F(\xi) = \sqrt{ch(\xi)/2}\,\beta_{01}^{-1}$. Using this function, we rewrite equation (3.27) as

$$F'' + \frac{1}{4}\left[\frac{3}{ch^2(\xi)} - 1\right]F = \frac{1}{F^3}. \qquad (3.30)$$

Solution of equation (3.30) reads as[7]

$$F(\xi) = \sqrt{\sum_{r,t=1}^{2} a_{rt} f_r(\xi) f_t(\xi)},$$

where $(f_1, f_2)$ is fundamental solution system of the equation

$$f'' + \frac{1}{4}\left[\frac{3}{ch^2(\xi)} - 1\right]f = 0, \qquad (3.31)$$

where $\|a_{rt}\|$ is symmetric matrix which satisfies the normalized condition

$$\det\|a_{rt}\| \cdot [W(f_1, f_2)]^2 = 1, \qquad (3.32)$$

where Wronskian $W(f_1, f_2) = f_1 f_2' - f_1' f_2$. By introducing new variable $t=\text{th}(\xi)$, equation (3.31) is reduced to associated Legendre differential equation[19]. The solution of the considered equation is

$$f_1(\xi) = 1/\sqrt{ch(\xi)},$$
$$f_2(\xi) = sh(\xi)/\sqrt{ch(\xi)}. \qquad (3.33)$$

Substituting functions (3.33) into the normalized condition (3.32), we define the elements of matrix $\|a_{rt}\|$ as

$a_{11}=a_{22}=1$, $a_{12}=a_{21}=0$. Therefore, the solution of equation (3.30) is

$$F(\xi) = \sqrt{ch(\xi)} \qquad (3.34)$$

Using the definition of function $F(\xi)$ and formula (3.34), we obtain

$$\beta_{01} = \sqrt{1/2} \qquad (3.35)$$

Substituting formula (3.35) into expression (3.26) for $\alpha_{02}$ we get

$$\alpha_{02}=0. \qquad (3.36)$$

Taking into account formulas (3.35) and (3.36), equation (3.28) for unknown function $\beta_{00}(\xi)$ can be written in the form

$$ch(\xi)[ch(\xi)\beta_{00}']' + \beta_{00} = 0. \qquad (3.37)$$

Solution of equation (3.37) can be written as[7]

$$\beta_{00}(\xi) = \frac{f(\xi)}{F(\xi)} = B_1 \frac{1}{ch(\xi)} + B_2 th(\xi), \tag{3.38}$$

where $B_1$ and $B_2$ are arbitrary constants defined from the boundary conditions. From expression (3.26) and (3.38), we obtain that

$$\alpha_{01} = B_1 th(\xi) - B_2 \frac{1}{ch(\xi)}.$$

Function $\alpha_{00}(\xi)$ is defined from (3.29) using expressions (3.35), (3.36) and (3.38).

$$\alpha_{00}(\xi) = \frac{i}{2} \ln[ch(t)]\Big|_{d_0}^{\xi} \pm \frac{B_{1,2}}{2} \frac{th(t)}{ch(\xi)}\Big|_{d_0}^{\xi} \mp \left(q + \frac{1}{2}\right) arctg[sh(\xi)] \tag{3.39}$$

The next pair of equation (3.16) and (3.17) at $m=1$ reads

$$a_1(\xi,v) = -ch(\xi)\frac{\partial \alpha_1}{\partial \xi} + i\frac{\partial^2 \alpha_1}{\partial v^2} - 2\left(\frac{\partial \alpha_0}{\partial v}\right)\left(\frac{\partial \alpha_1}{\partial v}\right) + \sqrt{2}\beta_0^2 \frac{\partial \beta_1}{\partial v} + \beta_0 \beta_1 - \sqrt{2}(2q+1)\frac{\partial \beta_1}{\partial v}$$

$$+ \frac{c^2}{2L_B^2}\left[v \cdot ch(\xi) - sh(2\xi)\frac{\partial \alpha_0}{\partial v}\right] = 0$$

$$b_1(\xi,v) = ich(\xi)\frac{\partial \beta_1}{\partial \xi} + \frac{\partial^2 \beta_1}{\partial v^2} + i2\left(\frac{\partial \alpha_0}{\partial v}\right)\left(\frac{\partial \beta_1}{\partial v}\right) + i\sqrt{2}\frac{\partial \alpha_1}{\partial v} + i\frac{1}{\sqrt{2}}\frac{c^2}{2L_B^2} sh(2\xi) = 0. \tag{3.40}$$

From equations (3.40), it follows that $\alpha_1(\xi,v)$ is polynomial of the first order in $v$ and $\beta_1(\xi,v)$ is polynomial of the zero order.

$$\alpha_1(\xi,v) = \alpha_{11}(\xi)\, v + \alpha_{10}(\xi),$$
$$\beta_1(\xi,v) \equiv \beta_{10}(\xi). \tag{3.41}$$

Substituting pair of polynomials (3.41) into system of equations (3.40) we obtain

$$\left[-ch(\xi)\alpha'_{11} + \frac{\beta_{10}}{\sqrt{2}} + \frac{c^2}{2L_B^2} ch(\xi)\right] \cdot v + \left[-ch(\xi)\alpha'_{10} - 2\alpha_{01}\alpha_{11} + \beta_{10}\beta_{00} - \frac{c^2}{2L_B^2} sh(2\xi)\alpha_{01}\right] = 0$$
$$\tag{3.42}$$

$$ch(\xi)\beta'_{10} + \sqrt{2}\alpha_{11} + \frac{1}{\sqrt{2}}\frac{c^2}{2L_B^2} sh(2\xi) = 0. \tag{3.43}$$

We express $\alpha_{11}$ from equation (3.43)

$$\alpha_{11} = -\frac{ch(\xi)}{\sqrt{2}}\beta'_{10} - \frac{1}{4}\frac{c^2}{L_B^2} sh(2\xi). \tag{3.44}$$

Formula (3.44) is substituted into (3.42). We obtain an equation for $\beta_{10}(\xi)$ from the condition that the coefficients in front of $v$ are equal to zero in formula (3.42)

$$ch(\xi)[ch(\xi)\beta'_{10}]' + \beta_{10} = -\sqrt{2}\frac{c^2}{L_B^2}ch^3(\xi). \tag{3.45}$$

Equating the left-hand side of equation (3.45) to zero, we get the inhomogeneous equation (3.37). So, the general solution of equation (3.45) is

$$\beta_{10}(\xi) = \beta_{00}(\xi) - \frac{1}{\sqrt{2}}\frac{c^2}{L_B^2}ch(\xi).$$

Solution (3.13) tends to zero when $v \to \pm\infty$ if index $q$ is positive integer, i.e., $q=0, 1, 2…$ In this case, function of parabolic cylinder can be expressed in terms of the Hermit polynomial[19] such that $H_q(-x)=(-1)^q H_q(x)$. Boundary condition (3.3) is satisfied with accuracy $O\left(\omega^{-1/2}\right)$ when $B_1=B_2=0$. Therefore, we obtain

$$\beta_0(v) = \pm v/\sqrt{2},$$

$$\alpha_{01}=0,$$

$$\beta_1(\xi) = -\frac{1}{\sqrt{2}}\frac{c^2}{L_B^2}ch(\xi),$$

$$\alpha_1 = const. \tag{3.46}$$

Taking into consideration formulas (3.19), (3.23), (3.35), (3.36), (3.39), and (3.46), solution (3.13) can be written in the first order approximation as

$$U_{\pm}(\xi,v) = \frac{1}{\sqrt{ch(\xi)}}\exp\left\{\pm\frac{i\omega}{2}sh(\xi)\mp i\left(q+\frac{1}{2}\right)arctg[sh(\xi)]\right\}D_q\left[v - \omega^{-1/2}\frac{c^2}{L_B^2}ch(\xi)\right]. \tag{3.47}$$

Solution (3.47) satisfies boundary condition (3.2). Repeating the calculation given in subsection 3.A, we obtain the following quantization condition for the energy spectrum

$$\omega_{p,q} = \frac{1}{sh(\bar{\xi})}\left\{\pi p + (2q+1)arctg[sh(\bar{\xi})]\right\}, \tag{3.48}$$

where $p$ is integer number such that $p \gg q$. Formula (3.48) can obtained from (3.9) when hyperbolic caustic coordinate $\varphi_{hc}$ is rather small. At the neighborhood of the least axis of the boundary ellipse, the general asymptotic solution of equation (3.1) is the following

$$U_{p,q}(\xi,v) = \frac{const}{\sqrt{ch(\xi)}}\frac{\sin}{\cos}\left\{\frac{\omega_{p,q}}{2}sh(\xi) - \left(q+\frac{1}{2}\right)arctg[sh(\xi)]\right\}D_q\left[v - \omega_{p,q}^{-1/2}\frac{c^2}{L_B^2}ch(\xi)\right].$$
$$\tag{3.49}$$

sin (cos) is taken when p is even (odd) in formula (3.49). Function of parabolic cylinder $D_q\left(\sqrt{2}Z\right)$ can be expressed in terms of Hermit polynomial $H_q(Z)$ when index $q$ is integer[19]

$$D_q\left(\sqrt{2}Z\right) = \exp\left\{-Z^2/2\right\}H_q(Z)$$

Formulas (3.48) and (3.49) coincide to those obtained by Buldyrev[8] in the magnetic field absence.

Function $D_q(\sqrt{2}Z)$ exponentially decreases when $|Z| > \sqrt{2q+1}$. Hence, function $U_{p,q}(\xi,\nu)$ oscillates in the band defined by the inequality

$$\sin(\varphi) \leq \pm \omega_{p,q}^{-1/2} \sqrt{2q+1} + \omega_{p,q}^{-1} \frac{c^2}{L_B^2} ch(\xi). \qquad (3.50)$$

The equation in expression (3.50) gives the value of the hyperbolic caustic coordinate $\varphi_{hc}$ of the right (left) hyperbola branch when sign plus (minus) is taken in front of the radical. Formula (3.50) is satisfied at large values of number $p$. In the first order of approximation, the eigenvalue is not dependent on the magnetic field for the harmonic oscillator states. Nevertheless, the magnetic field affects the hyperbolic caustic. This effect is not uniform. It increases towards the ellipse boundary.

## 4. NUMERICAL CALCULATION

We consider the energy spectrum of the holes confined in the bismuth circular wire when the energy lies in the vicinity of the valence band T-point. The eigenvalue of equation (2.1) is proportional to the square root of the absolute value of the hole energy measured from the top of the bismuth T-valence band. In the rombohedral system of coordinates, the bismuth wire is grown along direction [10-11]. In the wire cross section, the values of the effective mass components of the T-hole pocket are $m_1=0.0590\ m_o$, $m_2=0.3261\ m_o$[1]. In this case, the mass anisotropy is rather large. Hence, in the equivalent isotropic effective mass model for an elliptic potential well, the boundary is represented as a rather elongated ellipse. The half of the distance between the boundary ellipse focuses is $c=693$ nm when the wire radius is R=500 nm. Parameter $c$ is less than the wire radius when the mass anisotropy is small. In this case, the ellipse eccentricity (measure of the ellipse elongation) is $e=0.9$.

Let's consider dependence of the eigenvalue of the boundary states (BS) on quantum numbers $q$ and $p$ as well as on applied magnetic field. Figure 1(a) presents dependence of the dimensionless eigenvalue, $\omega(p,q,\lambda)$, on angular momentum quantum number $q$ at different values of $p$. Radial quantum number, $p$, corresponding to the boundary state subband is taken to be equal to 1, 2, and 3. Quantum number $q$ corresponds to different modes of the BS subband. Number $\lambda$ is equal to +1 and -1. It indicates the parallel and antiparallel direction of the magnetic field along the wire. The magnetic length, $L_B$, is supposed to be twice the wire radius $R$. Figure 1 shows that the step height of the eigenvalue staircase decreases when quantum number $q$ increases. For example, $\omega(1,2,+1)-\omega(1,20,+1)=2.61$, $\omega(1,121,+1)-\omega(1,120,+1)=2.50$. The split of the energy levels provided by the applied magnetic field slightly decreases (increases) when quantum number $q$ ($p$) increases. For instance, $\omega(1,20,+1) - \omega(1, 20, -1)=2.62$, $\omega(1,120,+1)-\omega(1,120,-1)=2.50$, $\omega(2,120,+1)-\omega(2,120,-1)=2.54$. Acting in opposite directions, the magnetic field symmetrically shifts the energy level with respect to the unperturbed level position. For example, $\omega(1,120,+1)-\omega(1,120,0)=1.25$ and $\omega(1,120,0)-\omega(1,120,-1)=1.25$, where $\omega(1,120,0)$ is the energy value at zero magnetic field ($\lambda=0$). The split between the eigenvalues due to the applied magnetic field leads to the following relation $\omega(p,q+1,-1)=\omega(p,q,+1)$. The difference between the eigenvalues corresponding to different values of $p$ increases when $q$ increases, that is, the distance between the subband edges increases. For example, $\omega(2,20,+1)-\omega(1,20,+1)=9.29$, $\omega(3,20,+1)-\omega(2,20,+1)=7.95$, $\omega(2,120,+1)-\omega(1,120,+1)=15.8$, $\omega(3,120,+1)-\omega(2,120,+1)=13.1$. The difference between the eigenvalues decreases at fixed value of $q$ and increasing $p$. Therefore, the density of boundary states tends to increase when $p$ increasing and vise versa when $q$ increasing.

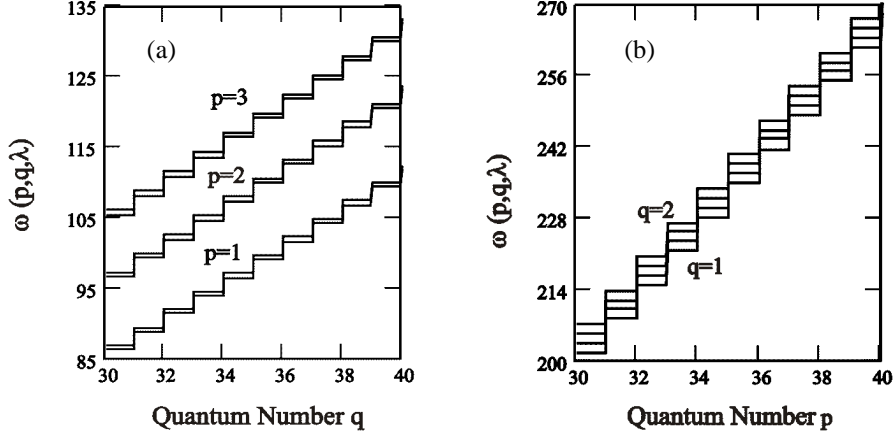

Fig. 1. (a) Dependence of the BS hole dimensionless eigenvalue on the angular momentum quantum umber, $q$, for two opposite directions of magnetic field, $\lambda=+1, -1$ when the radial quantum numbers $p=1, 2, 3$. (b) Dependence of the even and odd HOS dimensionless eigenvalue on the radial quantum number, $p$, at $q=1, 2$. Bismuth wire radius R=500 nm.

At given values of the hole mass components, the wire boundary coordinate is $\xi_{bound}=0.454$ in the elliptic system of coordinates. The elliptic coordinate is $\xi_{ec}=0.240$ and $\xi_{ec}=0.3704$ when the set of quantum numbers $\{p,q\}$ is correspondingly given by $\{1,30\}$ and $\{1,120\}$ at one direction of the magnetic field. The respective eigenvalues are $\omega(1,30,+1)=87.7$ and $\omega(1,120,+1)=315$. If the direction of the magnetic field is changed, then the elliptic coordinate is $\xi_{ec}=0.238$ and $\xi_{ec}=0.3703$, correspondingly. In this case, the eigenvalues are $\omega(1,30,+1)=85.1$ and $\omega(1,120,+1)=312$. The split between the caustic coordinates due to the applied magnetic field decreases when quantum number $q$ increases. In the presence of the magnetic field, the following relation $\xi_{ec}(p,q+1,-1)=\xi_{ec}(p,q,+1)$ is valid. At fixed radial quantum number $p$, the particle comes near to the boundary when the magnetic quantum number, $q$, increases. The applied magnetic field pushes the particle towards the boundary for one field direction and repulses it for the opposite direction. For example, the elliptic caustic coordinate is $\xi_{ec}=0.240$, $\xi_{ec}=0.239$, and $\xi_{ec}=0.238$ at $\lambda=+1$, $\lambda=0$, and $\lambda=-1$ ($p=1$, $q=30$), respectively.

In the rectangular system of coordinates, the equation of the elliptic caustic is given by equation

$$\frac{x^2}{a^2}+\frac{y^2}{b^2}=1.$$

The half of the longitudinal and transverse axis is $a=465.56$ nm and $b=257.3$ nm ($a/b=1.809$, eccentricity $e=0.833$) when $q=30$, $p=1$, $\lambda=+1$, $a=484$ nm, and $b=403$ nm ($a/b=1.2$, eccentricity $e=0.55$) when $q=120$, $p=1$, and $\lambda=+1$. The ratio of the longitudinal and transverse axis and eccentricity, i.e., the caustic anisotropy factor, decreases and the elliptic caustic tends to the circular wire boundary when the magnetic quantum number increases. The half of the longitudinal and transverse axis is $a=465.42$ nm and $b=255.9$ nm ($a/b=1.82$, eccentricity $e=0.835$) when $q=30$, $p=1$, $\lambda=-1$, $a=465.49$ nm, and $b=256.6$ nm ($a/b=1.814$, eccentricity $e=0.834$) when $q=30$, $p=1$, and $\lambda=0$. Hence, the caustic anisotropy factor increases or decreases in dependence on direction of the magnetic field. When the radial quantum number increases, the boundary ring becomes wider at fixed angular momentum quantum number.

Our calculations show that the split between caustics corresponding to opposite directions of the magnetic field exponentially decays with increasing ratio $L_B/R$. The elliptic caustic does not intersect the boundary under an applied magnetic field in quasi-classical limit. For example, $\xi_{ec}=0.248$ and $\xi_{ec}=0.371$ at $q=30$ and $q=120$, respectively, ($L_B/R=1$, $p=1$, $\lambda=+1$), whereas $\xi_{bound}=0.454$. Hence, the idea of Bogachek[16] that the oscillations of the physical magnitudes in the magnetic field are due to the intersections of the caustic with the cylindrical wire boundary is not valid. The numerical calculation shows that the elliptic caustic varies linearly with increasing of the magnetic field.

The dependence of the eigenvalue on quantum numbers and space variation of the caustic coordinate are mainly similar for both the BS and RS. The difference is the following. For the ring states the eigenvalue is less and the localization domain is greater in comparison with those for the Boundary State at the same quantum numbers $q$ and $p$. For example, $\omega(1,30,+1)=86.4$, $\xi_{ec}(1,30,+1)=0.235$ for the ring state when $q=30$ and $p=1$. Relations $\omega(p,q+1,-1)=\omega(p,q,+1)$ and $\xi_{ec}(p,q+1,-1)=\xi_{ec}(p,q,+1)$ are not valid for the RS. For instance, $\omega(1,31,-1)=88.7$, $\omega(1,30,+1)=86.4$, $\xi_{ec}(1,31,-1)=0.239$, and $\xi_{ec}(1,30,1)=0.235$. The reason of this difference is that the term proportional to the magnetic field in the dispersion relation (2.42) includes the caustic coordinate that depends on quantum numbers.

Figure 1 (b) shows the dependence of the eigenvalue, $\omega(p,q)$, for the hyperbolic caustic state (HCS) on the quantum number, $p$, at different values of $q$. Quantum number $q$ corresponding to HCS subband is taken to be equal to 1, 2. There are two different states corresponding to the same quantum number $q$. One of the states corresponds to the even number of oscillations of the wave function along $X$ axis. The other state corresponds to the odd number of the oscillations. Quantum number $p$ corresponds to different modes of the HCS subband. In our approximation, the HCS energy does not depend on magnetic field. The distance between the energy levels corresponding to the different modes of the same HCS subband is conserved with increasing of quantum number $p$. For example, $\omega_{even}(31,1)-\omega_{even}(30,1)=6.68$ and $\omega_{even}(111,1)-\omega_{even}(110,1)=6.68$. The distance between the different band edges is also conserved. For instance, $\omega_{even}(40,2)-\omega_{even}(40,1)=3.64$, $\omega_{even}(120,2)-\omega_{even}(120,1)=3.64$, $\omega_{even}(40,2)-\omega_{odd}(40,1)=5.61$, and $\omega_{even}(120,2)-\omega_{odd}(120,1)=5.61$. Hence, the density of the hyperbolic caustic states is constant. For corresponding subbands, the eigenvalue for the hyperbolic caustic state is about in three times greater than that for the ring state at the similar set of quantum numbers $q$ and $p$ ({1,30} and {30,1}). The energy levels (with similar numbers $p$ and $q$) of corresponding subbands coincide for the hyperbolic caustic and harmonic oscillator states (HOS) with exception of the odd HCS subband when $q=1$. There is no HOS subband which corresponds to the 1st odd HCS subband.

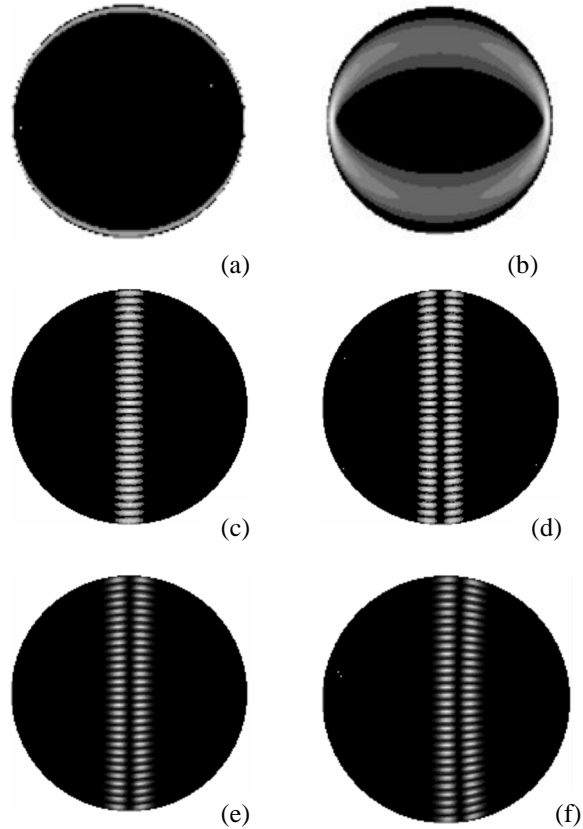

Fig. 2 Counter plot of the T hole wave function corresponding to
(a) BS, (b) RS, (c) odd HCS, (d) even HCS, (e) HOS at $L_B=2$ R,
and (f) HOS at $L_B=0.3$ R. Bismuth wire radius R=500 nm.

Figures 2 (a-b) depict the density distribution of the wave function corresponding to the BS and RS in the wire cross section at $q=30$ and $p=1$. The holes are tightly localized in the vicinity of the boundary for BS. This agrees with the result presented by J. C. Gutlerrez-Vega et al.[20] The wave function oscillates in such a way that the oscillation amplitude decays towards the caustic in the boundary ring. For the RS, particles greater localize near the elliptic caustic than near the boundary.

Figures 2 (c-d) depict the particle space probability distribution for the hyperbolic caustic states in the wire cross section at $p=30$ and $q=1$. In the isotropic mass model, the holes are localized at the vicinity of the ellipse minimal diameter. The number of eigenfunction oscillations is equal to 30 along $Y$ axis as well as 1 (odd state) and 2 (even state) along $X$ axis.

When the set of hyperbolic caustic state (HCS) quantum numbers $\{p,q\}$ is given by $\{30,1\}$ and $\{120,1\}$, the hyperbolic coordinate is $\varphi_{hc}=0.092$ rad. ($\approx 5°$) and $\varphi_{hc}=0.0461$ rad. ($\approx 3°$), correspondingly. The respective eigenvalues are $\omega_{odd}(30,1)=201$ and $\omega_{odd}(120,1)=803$. The hyperbolic coordinate is $\varphi_{hc}=0.11$ rad. ($\approx 6°$) when the set of quantum numbers $\{p,q\}$ is given by $\{120,2\}$. The respective eigenvalues are $\omega_{odd}(120,1)=803$ and $\omega_{odd}(120,2)=807$. At the fixed angular momentum quantum number, $q$, the particle localization domain between two hyperbolic caustics is narrowed when radial quantum number $p$ increases. Hence, in contrast to the BS and RS, the increase of the radial quantum number, $p$, rises an electron (hole) localization for HCS.

Figure 2 (e) shows the space variation of the probability density for the harmonic oscillator state hole eigenfunction related to the 1$^{st}$ ($q=1$) HOS subband in the wire cross section when $p=30$. It corresponds to the 1$^{st}$ even HCS subband. The 2$^{nd}$ HOS subband corresponds to the 2$^{nd}$ odd HCS subband. Since HOS hyperbolic caustic coordinates are less than the respective HCS caustics, the HOS are greater localized in space than HCS. For example, averaged value $\varphi_{hc}^{HOS} = 0.124$ rad. ($\approx 7°$) and exact $\varphi_{hc}^{HCS} = 0.171$ rad. ($\approx 10°$) corresponding to common eigenvalue $\omega(30,1)=203$. If the value of the eigenvalue is $\omega(30,2)=205$, then the value of the hyperbolic caustic coordinate is $\varphi_{hc}^{HOS} = 0.159$ rad. ($\approx 9°$) (averaged) and $\varphi_{hc}^{HCS} = 0.218$ rad. ($\approx 12°$) (exact). The magnetic field inflects the HOS caustics from their initial position. For instance, the hyperbolic caustic coordinate is $\varphi_{hc}^{HOS} = \pm 6.98°$ in the absence of the magnetic field ($p=30$, $q=1$). The right hand branch hyperbolic caustic coordinate is $\varphi_{hc}^{HOS}(\xi_{bound}) = 7.13°$ at the boundary and $\varphi_{hc}^{HOS}(\xi = 0) = 7.11°$ at the intersection with $X$ axis. The left hand branch hyperbolic caustic coordinate is $\varphi_{hc}^{HOS}(\xi_{bound}) = -6.83°$ at the boundary and $\varphi_{hc}^{HOS}(\xi = 0) = -6.84°$ at the intersection with $X$ axis when $L_B/R=2$. Therefore, the influence of the applied magnetic field on the hyperbolic caustics is not uniform along the whole length of the caustic. The number of HOS eigenfunction oscillations along the $X$ axis is equal to number $q$ of the corresponding HOS subband. The number of the oscillations along the $Y$ axis is equal to number $p$ of the corresponding HOS subband mode. As it was noted above, there is no HOS subband which corresponds to the 1$^{st}$ odd HCS subband; therefore, the number of the HOS eigenfunction oscillations along the $X$ axis is always greater than 1.

To visually observe the effect of the magnetic field on the HOS hyperbolic caustics we have to fall outside the limits of our approximation. Figure 2 (f) depicts probability density distribution of the hole HOS eigenfunction when $L_B/R$=0.3. If the direction of the magnetic field is changed to opposite one, then the plot is symmetrically reflected with respect to the *Y* axis. Our calculations show that the electron (hole) localization and corresponding eigenvalues of the BS, RS, HCS, and HOS are different. So, they should be treated separately in the semi-classical approximation.

## 5. Conclusions

In a weak longitudinal magnetic field, the asymptotic solution of the Schrodinger equation for a particle confined by the impenetrable elliptic wire boundary gives four different excited states. There are boundary state (BS) and ring state (RS) grouped together into "whispering gallery" modes. There are hyperbolic caustic state (HCS) and harmonic oscillator state (HOS) which are grouped together into "jumping ball" modes. The radial quantum number, *p*, corresponds to the BS (RS) subband, the angular momentum quantum number, *q*, does to the respective subband mode. The angular momentum quantum number is much greater than the radial quantum number for these states. As contrasted, the angular momentum quantum number, *q*, corresponds to HCS (HOS) subband, radial quantum number *p* does to respective HCS (HOS) subband mode. In contradistinction to BS and RS, the angular momentum quantum number is much less than the radial quantum number for HCS and HOS.

The "whispering gallery" mode corresponds to the particle (wave) moving in the ring restricted by the ellipse boundary and elliptic caustic. Its eigenfunction presents the exponent function multiplied by the Airy function. The arguments of these functions are series in terms of $E^{-1/6}$ (*E*- energy). The eigenfunction oscillates in the ellipse boundary layer and exponentially decays from the elliptic caustic to the ellipse focuses. The boundary ring is much less for the BS than for the RS.

The "jumping ball" mode corresponds to the particle (wave) moving in the band restricted by the boundary and two branches of the hyperbolic caustic. The respective eigenfunction oscillates between the branches of the hyperbolic caustics and exponentially decays from the hyperbolic caustic to the ellipse focuses. As well as the BS and RS eigenfunctions, the HCS eigenfunction presents the exponent function multiplied by the Airy function. In contrast, the HOS eigenfunction presents the exponent function multiplied by the parabolic cylinder function. The argument of the HOS function is series in terms of $E^{-1/4}$.

For the boundary and ring state, the density of states is enhanced but the particle localization is diminished when the radial quantum number, *p*, increases. If the angular momentum quantum number, *q*, increases, then both the density of states and electron (hole) localization rise. The density of hyperbolic caustic states is constant. In the first order approximation, both the HCS eigevalues and hyperbolic caustic are independent on magnetic field. For the harmonic oscillator and hyperbolic caustic states, the particle localization enhances when the radial quantum number, $p$, increases. In contradistinction to ring states, this localization diminishes when the angular momentum quantum number, $q$, increases.

The applied magnetic field splits caustics and eigenvalues of the BS and RS. This split decreases when the quantum number, $q$, increases. Due to applied magnetic field, the split between the BS eigenvalues, $\omega$, and caustics, $\bar{\xi}$, leads to the following relations $\omega(p,q+1,-1)=\omega(p,q,+1)$ and $\xi_{ec}(p,q+1,-1)=\xi_{ec}(p,q,+1)$. For the ring states, these relations are not valid. The effect of the magnetic field on the elliptic caustic is uniform along whole length of the caustic.

There are even and odd hyperbolic caustic states corresponding to the same angular momentum quantum number, $q$. The even and odd HCS are defined from the different dispersion relations. They associate with $2q$ and $2q$-1 eigenfunction oscillations along the $X$ axis. The number of eigenfunction oscillations along the $Y$ axis is given by quantum number $p$.

For the harmonic oscillator states, the eigenvalue is not dependent on weak magnetic field. Nevertheless, in contradistinction to the hyperbolic caustic state, the magnetic field affects the hyperbolic caustic. This effect increases near the wire boundary. Hence, it is not uniform along whole length of the hyperbolic caustic. There is no HOS subband which corresponds to the 1st odd HCS subband, therefore, the number of the HOS eigenfunction oscillations along the X axis is always greater than 1.